\documentclass[amsmath,amssymb,prb,floatfix,showpacs]{revtex4}
\newcommand{\figwid}{0mm}

\newcommand{\booo}{BO$_3$}
\newcommand{\cm}{cm$^{-1}$}
\newcommand{\rfb}{RFe$_3$(BO$_3$)$_4$}
\newcommand{\nfb}{NdFe$_3$(BO$_3$)$_4$}
\newcommand{\gfb}{GdFe$_3$(BO$_3$)$_4$}
\newcommand{\tfb}{TbFe$_3$(BO$_3$)$_4$}
\newcommand{\efb}{ErFe$_3$(BO$_3$)$_4$}
\newcommand{\yfb}{YFe$_3$(BO$_3$)$_4$}

\usepackage{graphicx}
\usepackage{dcolumn}
\usepackage{bm}
\renewcommand{\includegraphics}{\relax}

\begin{document}

\preprint{APS/123-QED}

\title{Raman scattering from phonons and magnons in RFe$_{3}$(BO$_{3}$)$_{4}$}
\author{Daniele Fausti}
\email{d.fausti@rug.nl}
\author{Agung Nugroho}
\author{Paul H.M. van Loosdrecht}
\email{P.H.M.van.Loosdrecht@rug.nl}

\affiliation{Material Science Centre, University of Groningen,
9747 AG Groningen, The Netherlands.}

\author{Sergei A.Klimin}
\author{Marina N. Popova}
\affiliation{Institute of Spectroscopy, RAS, 142190, Troitsk,
Moscow Region, Russia.}

\author{Leonard N. Bezmaternykh}
\affiliation{L.V. Kirensky Institute of Physics, Siberian Branch
of RAS, Krasnoyarsk, 660036, Russia.}

\date{\today}

\begin{abstract}
Inelastic light scattering spectra of several members of the
RFe$_{3}$(BO$_{3})_{4}$ family reveal a cascade of phase
transitions as a function of temperature, starting with a
structural, weakly first order, phase transition followed by two
magnetic phase transitions. Those consist of the ordering of the
Fe-spin sublattice revealed by all the compound, and a subsequent
spin-reorientational transition for GdFe$_{3}$(BO$_{3})_{4}$. The
Raman data evidence a strong coupling between the lattice and
magnetic degrees of freedom in these borates. The Fe-sublattice
ordering leads to a strong suppression of the low energy magnetic
scattering, and a multiple peaked two-magnon scattering continuum
is observed. Evidence for short-range correlations is found in the
`paramagnetic' phase by the observation of a broad magnetic
continuum in the Raman data, which persists up to surprisingly
high temperatures.

\end{abstract}


\maketitle
\section{Introduction}
The family of RM$_{3}$(BO$_{3})_{4}$ crystals (R=Y or a rare
earth, M=Al, Ga, Sc, Cr, Fe) has attracted considerable attention
during the last years. Different combinations of R and M lead to a
large variety of physical properties that, together with their
excellent physical characteristics and chemical stability, make
these crystals extremely interesting both from application and
fundamental points of view. The lack of inversion symmetry (they
crystallize in a trigonal space group) has stimulated different
applications in the field of optical and optoelectronic devices.
Crystals of YAl$_{3}$(BO$_{3})_{4}$ and GdAl$_{3}$(BO$_{3})_{4}$
doped with Nd have been widely studied and used in optical
devices, for self-frequency doubling and self-frequency summing
lasers \cite{jaq01,bre04,che01}. Concentrated
NdAl$_{3}$(BO$_{3})_{4}$ crystals are efficient media for
mini-lasers \cite{che01}. Rare-earth borates with magnetic ions (M
= Cr, Fe) are much less studied. Recently, some iron borates were
reported to possess multiferroic features, demonstrating the
coexistence of magnetic and ferroelectric order parameters
\cite{zve05}; they may therefore be considered for optoelectronic
applications. These magnetic borates are also expected to exhibit
interesting magnetic properties because of the presence of two
different kinds of magnetic ions (3d and 4f elements), and in
particular because their structure features isolated helicoidal
chains of magnetic 3d atoms.

\begin{figure}
\includegraphics[width=\figwid]{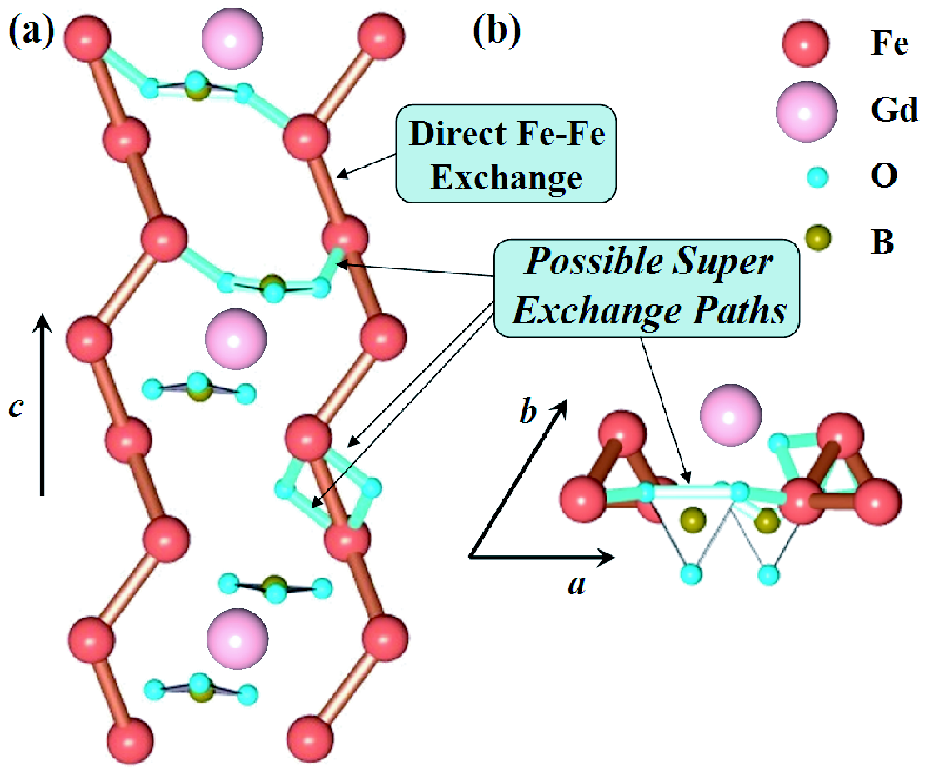}
\caption{\label{fig1} (Color online) The structure of
RM$_{3}$(BO$_{3}$)$_{4}$ incorporates helicoidal chains of
M$^{3+}$ ions. The picture shows the structure of iron borates
(M=Fe) viewed from different angles a) perpendicular to the
C$_{3}$ axis and b) parallel to it. The possible exchange paths
are shown.}
\end{figure}

The crystal structure of RM$_{3}$(BO$_{3})_{4}$ borates belongs to
the structural type of the natural mineral huntite
CaMg$_{3}$(CO$_{3})_{4}$ that crystallizes in the space group
$R$32 of the trigonal system \cite{jou68,bel79,cam97,kli05}. The
primitive unit cell contains one formula unit. Three kinds of
coordination polyhedra are present, trigonal prisms for RO$_{6}$,
octahedra for MO$_{6}$, and two types of planar triangular
BO$_{3}$ groups: equilateral B1O$_{3}$ and isosceles B2O$_{3}$
(the numbering of inequivalent ions position is the same as in
Ref. \cite{kli05}). The MO$_{6}$ octahedra share edges forming
mutually independent helicoidal chains which run parallel to the
$c$-axis (see Fig.\ref{fig1}). The RO$_{6}$ prisms are isolated
polyhedra, each of them connects three helicoidal MO$_{6}$ chains,
while there are no direct R-O-R links between different RO$_{6}$
prisms. This structure can also be seen as a succession of
alternating BO$_{3}$ and R+M layers perpendicular to the c axis.

Recently, a systematic study of rare-earth iron borates
RFe$_{3}$(BO$_{3})_{4}$ (R = Y, La-Nd, Sm-Ho) was undertaken for
polycrystalline samples \cite{hin03}. The measurements of
magnetization, specific heat, and $^{57}$Fe M\"{o}ssbauer spectra
revealed an antiferromagnetic ordering at low temperatures. The
magnetic ordering temperature increases with decreasing R$^{3 + }$
ionic radius, from 22~K for LaFe$_{3}$(BO$_{3})_{4}$ to 40~K for
TbFe$_{3}$(BO$_{3})_{4}$. In addition to this, X-ray diffraction,
specific heat measurements, and differential thermal analysis have
indicated a structural phase transition for
RFe$_{3}$(BO$_{3})_{4}$ compounds with R = Eu-Ho, Y. Its
temperature increases with decreasing R$^{3 + }$ ionic radius,
from 88~K for EuFe$_{3}$(BO$_{3})_{4}$ to 445~K for
YFe$_{3}$(BO$_{3})_{4}$. Magnetic ordering of the La- and Nd-iron
borates has been confirmed by spectroscopic and magnetic
measurements on non oriented single crystals \cite{cam97,chu04}.
Studies on oriented single crystals have only been carried out for
GdFe$_{3}$(BO$_{3})_{4}$, using a variety of methods
\cite{bal03,lev04,pan04,zve05,kli05,gav04}. Recent specific heat,
Raman scattering, and Nd$^{3 + }$ probe absorption measurements
revealed a cascade of three spontaneous phase transitions in
GdFe$_{3}$(BO$_{3})_{4}$ \cite{lev04}. A structural, weakly
first-order, phase transition into a less symmetric
low-temperature phase was found at $T_{S}$=156~K (this in contrast
to the powder result $T_{S}$=174 K \cite{hin03}). At lower
temperature, two magnetic transitions have been observed. A
second-order phase transition occurs at 37~K, resulting in an
antiferromagnetic ordering of the Fe spin sublattice. This
transition is accompanied by a polarization of the Gd spin
subsystem. The second magnetic transition is observed at 9~K, and
has been identified as a first-order Fe spin-reorientation
transition \cite{lev04}. A detailed picture of magnetic phase
transitions in GdFe$_{3}$(BO$_{3})_{4}$ as a function of
temperature and magnetic field has recently been suggested in Ref.
\cite{pan04} on the basis of antiferromagnetic resonance studies.

Quite recently, the low-temperature structural phase of
GdFe$_{3}$(BO$_{3})_{4}$ has been identified as a $P$3$_{1}$21
trigonal structure \cite{kli05}. To our knowledge, no information
exists on the low-temperature structure of other rare-earth iron
borates. Such information can be obtained from the comparison of
low-temperature Raman spectra of GdFe$_{3}$(BO$_{3})_{4}$ and
other compounds from the RFe$_{3}$(BO$_{3})_{4}$ family. Raman
studies of RFe$_{3}$(BO$_{3})_{4}$ were performed for R=La and Nd
at room temperature \cite{and97}. Preliminary temperature
dependent results on the R=Gd compound have also been reported
recently \cite{lev04}.

The present work investigates two different aspects of the
rare-earth iron borates. First, inelastic light scattering studies
of the vibrational and structural properties of different
RFe$_{3}$(BO$_{3})_{4}$ are presented. Amongst others they reveal
the weak first-order phase transition from the high-temperature
$R$32 structure to the low-temperature $P$3$_{1}$21 one. The
second part of this paper focuses on the magnetic properties.
Apart from observing changes in phonon frequencies at the
antiferromagnetic ordering transition arising from magneto-elastic
couplings, the spectra also exhibit relatively strong two-magnon
scattering features. The magnetically ordered phases exhibits well
defined magnon excitations resulting in a multiple peaked
scattering continuum at finite energy. In the paramagnetic phase a
strong low energy continuum is found, which results from the
persistence of short range correlations in the helicoidal Fe spin
chains up to quite high temperatures.

\section{EXPERIMENTAL DETAILS}
Different samples of RFe$_{3}$(BO$_{3})_{4}$, with R = Gd, Nd, Tb,
Er, and Y were grown using a Bi$_{2}$Mo$_{3}$O$_{12}$ - based flux,
as described in Ref. \cite{bezm1, bezm2}. Unlike the Bi$_{2}$O$_{3}$
based fluxes \cite{cam97}, in Bi$_{2}$Mo$_{3}$O$_{12}$ - based flux
Bi$_{2}$O$_{3}$ is strongly bonded to MoO$_{3}$ which excludes a
partial substitution of bismuth for a rare earth during
crystallization. Spontaneous nucleation from the flux resulted in
small single crystals (~1x1x1 mm). We used these crystals as seeds
to grow large (~10x5x5 mm) single crystals of Gd, Nd, and Tb iron
borates. As for ErFe$_{3}$(BO$_{3})_{4}$ and
YFe$_{3}$(BO$_{3})_{4}$, only small single crystals of these
compounds were studied in this work. All the crystals were green in
color. The samples were oriented either by X-ray diffraction
technique or by using their morphology and optical polarization
methods.

The Raman measurements were performed in a backscattering
configuration, using a three-grating micro-Raman spectrometer
(T64000 Jobin Yvon) equipped with a liquid nitrogen cooled charged
coupled device (CCD) detector. The frequency resolution was better
that 1~cm$^{ - 1}$ for the frequency region considered. The
samples were placed in a optical microscope cryostat. The
temperature was varied from 2.7 to 500~K, with a stability of $\pm
$0.02~K. The scattering was excited by the second harmonic light
of a Nd:YVO$_{4}$ laser (532~nm), focused down to $\sim $50~$\mu
$m$^{2}$ with the power density on the sample kept below
0.1~mW/$\mu $m$^{2}$. The polarization was controlled both on the
incoming and outgoing beams giving access to all relevant
polarization combinations.

Most of the experiments were performed with light incident
perpendicular to the C$_{3}$-axis of the crystal,
$\vec{k}\perp{c}$ (as a rule, precise orientation of $\vec {k}$ in
the (\textit{ab})-plane was not known), however, when the shape of
the samples allowed, also the $\vec{k}\vert \vert {c}$
configuration has been employed. Spectra are marked using the
Porto notation \cite{rou81}. For $\vec {k}\perp{c}$, only the
polarization of incident and scattered light are indicated. The
conventions z$\vert\vert $C$_{3}$, x$\vert \vert $C$_{2 }$ are
used to describe the geometry of the Raman experiments. We were
able to obtain only partially polarized Raman spectra of
ErFe$_{3}$(BO$_{3}$)$_{4}$ and YFe$_{3}$(BO$_{3}$)$_{4}$.

Preliminary dielectric measurements were performed for
RFe$_{3}$(BO$_{3})_{4}$ on non oriented crystals of typical size
1x1x0.2~mm$^{3}$. The surfaces were polished and covered with Ag
paste and the electrical contacts were made using Pt wires
connected to the surface by additional Ag paste. The samples were
mounted on a home-made insert for a Quantum Design PPMS system.
The capacitance was measured using a Andeen-Hagerling AH2500A
capacitance bridge operating at a fixed frequency of 1kHz.

\begin{figure}
\includegraphics[width=\figwid]{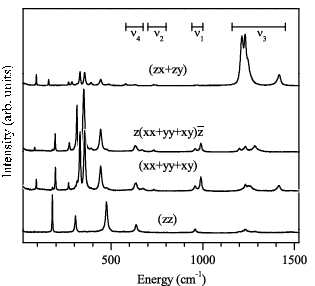}
\caption{\label{fig2} Polarized Raman spectra of
GdFe$_{3}$(BO$_{3}$)$_{4}$ at room temperature. The bars at the
top indicate the frequency regions for the internal vibrations of
BO$_{3}$ groups.}
\end{figure}
\begin{figure}
\includegraphics[width=\figwid]{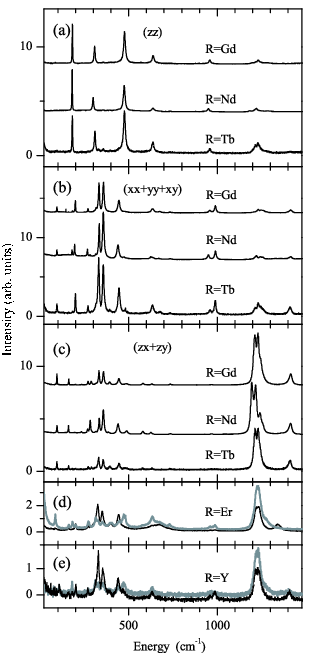}
\caption{\label{fig3} (a, b, c) Polarized Raman spectra for
RFe$_{3}$(BO$_{3}$)$_{4}$, R=Gd, Nd, Tb, at room temperature. (d,e)
Partially polarized Raman spectra for (d) ErFe$_{3}$(BO$_{3}$)$_{4}$
at 390 K and (e) YFe$_{3}$(BO$_{3}$)$_{4}$ at 420 K.}
\end{figure}

\section{EXPERIMENTAL RESULTS}
The room-temperature (RT) spectra of GdFe$_{3}$(BO$_{3})_{4}$ for
all studied experimental configurations are presented in
Fig.\ref{fig2}. The regions where the internal BO$_{3}$ vibrations
are expected are indicated by the horizontal bars in the top of
the panel. The spectra show a strong dependence on the
polarization of the incoming and scattered light, but also show a
distinct difference between (xx+yy+xy) polarized spectra for $\vec
{k}$ parallel to ${c}$ and for $\vec{k}$ perpendicular to ${c}$.
The spectra for different rare earth iron borates are, as
expected, qualitatively quite similar, as is shown in
Fig.\ref{fig3}.

\begin{figure}
\includegraphics[width=\figwid]{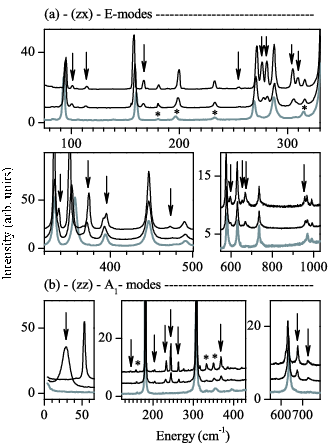}
\caption{\label{fig4} Raman spectra of GdFe$_{3}$(BO$_{3}$)$_{4}$
above and below $T_{s}$=156 K, namely, at 161 K (the lowest traces
in each panel), 150 K (middle traces), and 30 K (upper traces).
(a) (zx+zy) polarization; (b) (zz) polarization. New modes that
appear below $T_{s}$ are indicated by arrows. Asterisks mark the
lines from another polarization.}
\end{figure}

\begin{figure}
\includegraphics[width=\figwid]{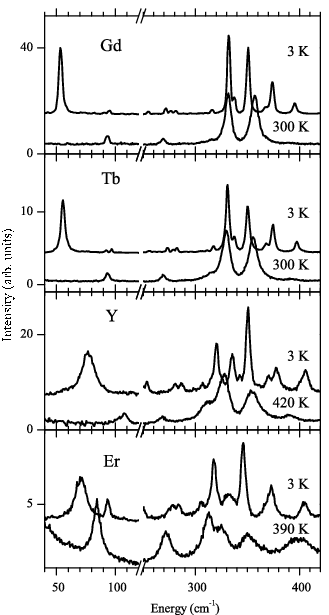}
\caption{\label{fig5} Comparison of several low-temperature new
modes in different RFe$_{3}$(BO$_{3}$)$_{4}$. Spectra in
(xx+xy)-polarization for R=Gd and Tb and partially polarized spectra
for R=Y and Er.}
\end{figure}

\begin{figure}
\includegraphics[width=\figwid]{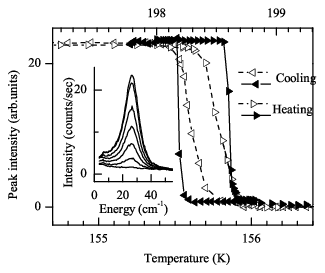}
\caption{\label{fig6} The lowest-frequency A$_{1}$-mode appearing in
low-temperature $P$3$_{1}$21 structure demonstrates hysteresis in a
plot of its intensity vs temperature. The data for Gd(Tb)-compound
are presented by open (black) triangles. Inset shows this mode
during cooling down for temperatures 154.91, 155.03, 155.08, 155.11,
155.16, 155.20, 155.45 K in Gd-compound. The lower temperature axis
refer to R=Gd, the upper one - to Tb.}
\end{figure}

The phonon lines narrow progressively with lowering the
temperature and several new modes appear abruptly in
GdFe$_{3}$(BO$_{3})_{4}$, TbFe$_{3}$(BO$_{3})_{4}$,
ErFe$_{3}$(BO$_{3})_{4}$, and ErFe$_{3}$(BO$_{3})_{4}$, at
$T_{S}$=155~K, 198~K, 340~K, and 350~K, respectively. Figures
\ref{fig4} and \ref{fig5} and Tab. \ref{tab3} give a detailed
comparative picture of the phonon modes observed above and below
$T_{S}$ in RFe$_{3}$(BO$_{3})_{4}$. The intensities of the new
modes show a hysteresis as a function of temperature (see
Fig.\ref{fig6}). The lowest-frequency intense new mode appears
abruptly below the phase transition, but then exhibits a
soft-mode-like behavior with further lowering of the temperature;
i.e. for \gfb~ and \tfb~ the low frequency mode shifts from
26~cm$^{ - 1}$ at $T_{S}$ to respectively 56~cm$^{ - 1}$ and
57~cm$^{ - 1}$ at 3~K, and for \efb~ and \yfb~ it goes
respectively from 25 to 72~cm$^{ - 1}$ and from 27 to 76~cm$^{ -
1}$. At the same time its linewidth decreases from 12 to 3~cm$^{ -
1}$ (see Figs.\ref{fig7} and 4b). The spectra in (xz+yz)
polarization of all the compounds demonstrate a low-frequency
scattering continuum that gradually changes its shape with
lowering the temperature and exhibits the opening of a ''gap'' at
$T_{N}$ (see Fig.\ref{fig8}). For NdFe$_{3}$(BO$_{3})_{4}$ at the
lowest temperatures a well-defined narrow peak at about 10~cm$^{ -
1}$ is observed within the gap. This mode shifts to zero frequency
and strongly broadens upon approaching $T_{N}$ from below
(Fig.\ref{fig8}(c)). For the Gd compound a similar mode is
observed below $T_{N}$ at about 18~cm$^{-1}$.

\section{GROUP-THEORETICAL ANALYSIS}
\subsection{High-temperature structure $R$32 ($D_3^7 $)}
The primitive cell of the high-temperature structure $R$32 of
RFe$_{3}$(BO$_{3})_{4}$ contains 20 atoms which results in 57
vibrational normal modes. Knowing the local symmetry of all the
atomic positions, one can perform the factor-group analysis to
find the symmetries of these modes \cite{mav80,rou81}. R and B1
atoms reside in highly symmetric $D_{3}$ positions
\cite{cam97,kli05} and generate $A_{2}\oplus E$ modes each. Fe,
B2, O1, and O2 atoms are in $C_{2}$ symmetry positions, which
results in $A_{1} \oplus $2$A_{2}\oplus$3$E$ modes for each of
them. O3 atoms occupy general $C_{1}$ positions and give 3$A_{1}
\oplus $3$A_{2} \oplus$6$E$ modes. Summing all these modes and
subtracting the $A_{2}\oplus E$ acoustic modes one gets the
following optical vibrational modes of the crystal, in accordance
with the results of Ref.\cite{and97}.

\begin{eqnarray}\label{eq1}
\Gamma _{vibr} = 7A_1 \mathrm{(xx,yy,zz)} \oplus 12A_2 (E //\mathrm{z})\nonumber\\
\oplus 19E(E / / \mathrm{x},E / / \mathrm{y,xz,yz,xy)}
\end{eqnarray}

Notations in parentheses refer to the allowed components of
electric dipole moment (infrared (IR) activity) and the
polarizability tensor (Raman activity). Doubly degenerated E modes
are polar and both IR and Raman active. The light propagating
along the $c$-axis ($\vec {k}\vert \vert c$) probes TO modes,
while the $\vec {k} \bot c$ configuration gives pure LO modes
\cite{and97}.

The strongest bonds in the structure of RFe$_{3}$(BO$_{3})_{4}$
borates are B-O bonds within the planar BO$_{3}$ triangles.
Considering them as molecular units that occupy $D_{3}$ (for
B1O$_{3})$ and $C_{2}$ (for B2O$_{3})$ positions, and adding the
modes of translational character coming from R and Fe ions
(respectively in $D_{3}$ and $C_{2}$ position), we obtain the
following optical modes:

\begin{equation}
\label{eq2} \Gamma _{vib}^{tr} = 2A_1 \oplus 5A_2 \oplus 7E; \quad
\Gamma _{vib}^{lib} = A_1 \oplus 3A_2 \oplus 4E,
\end{equation}
where the librational ones (right equation) come from the BO$_{3}$
units. The total number of external optical vibrations is, thus,

\begin{equation}
\label{eq3} \Gamma _{vib}^{ext} = 3A_1 \oplus 8A_2 \oplus 11E
\end{equation}

Subtracting (\ref{eq2}) from (\ref{eq1}), one gets the internal
vibrational modes related to the vibrations of BO$_{3}$ groups:

\begin{equation}
\label{eq4} \Gamma _{vib}^{int} = 4A_1 \oplus 4A_2 \oplus 8E
\end{equation}

The correlation scheme of Fig.\ref{fig9} helps to get a deeper
insight into the origin of the vibrational modes of
RFe$_{3}$(BO$_{3})_{4}$. It follows from this scheme that two of
the seven $A_1$ modes originate from the fully symmetric vibration
of the  B1O$_{3}$ and  B2O$_{3}$ groups, two correspond to the
B2O$_{3}$ $\nu _{3}$ and $\nu _{4}$ vibrations, and two to the
rotational and translational modes of the B2O$_{3}$ group. The
remaining $A_{1}$ mode results from displacements of the iron
atom. Table \ref{tab1} indicates the origin of crystal modes in
the frequency ranges of internal vibrations of the BO$_{3}$
molecules.

\subsection{Low-temperature structure $P$3$_{1}$21($D_3^4 $)}
The primitive cell of the low temperature structure contains three
formula units \cite{kli05} (60 atoms) which gives 177 vibrational
modes. Unique positions for the R and B1 atoms remain but their
symmetry lowers from $D_{3}$ in the $R$32 structure to $C_{2}$ in
the $P$3$_{1}$21 one. On the contrary, instead of single threefold
$C_{2}$ symmetry positions for Fe, B2, O1, and O2 there are now
two different positions, namely, a threefold $C_{2}$ and a sixfold
$C_{1}$ positions for each of them. O3 oxygen atoms occupy now
three different general ($C_{1})$ positions \cite{kli05}. The
factor-group analysis of the low-temperature structure
$P$3$_{1}$21 gives the following optical vibrational modes:

\begin{equation}
\label{eq5} \Gamma_{vib}(P3_121)=27A_1\oplus32A_2\oplus59E
\end{equation}

Carrying out the same procedure as in the previous subsection one
finds what modes correspond to the external (translational and
librational) motions of the R and Fe atoms and of the BO$_{3}$
entities and to the internal vibrations of the BO$_{3}$
"molecules". The results are summarized in Table \ref{tab2}.

\section{DISCUSSION}
\subsection{Phonons of RFe$_{3}$(BO$_{3}$)$_{4}$ in the $R$32 and $P$3$_{1}$21 phases}
First of all, we consider the polarized room-temperature Raman
spectra of RFe$_{3}$(BO$_{3})_{4}$, R=Nd, Gd, and Tb
(Fig.\ref{fig3}). Table \ref{tab3} summarizes the observed Raman
modes of these iron borates at RT. It has been constructed taking
into account the results of the group-theoretical analysis for the
$R$32 structure and the frequency ranges for normal modes of the
regular planar free ion (BO$_{3})^{3 - }$ (they are indicated in
Fig.\ref{fig2}; see also Tables \ref{tab1} and \ref{tab2}). The
spectra of ErFe$_{3}$(BO$_{3})_{4}$ and YFe$_{3}$(BO$_{3})_{4}$
were only partially polarized. Comparing the 390~K spectra of
ErFe$_{3}$(BO$_{3})_{4}$ and the 420~K spectra of
YFe$_{3}$(BO$_{3})_{4}$ with the RT spectra of other iron borates
we complete Table \ref{tab3} for these compounds as well (at room
temperature, ErFe$_{3}$(BO$_{3})_{4}$ and YFe$_{3}$(BO$_{3})_{4}$
are already in the low-temperature phase).

The low-frequency part of the Raman spectrum is dominated by
external modes arising from translational motions of the R and Fe
atoms and from librational and translational modes of BO$_{3}$
groups. The number and polarization properties of the modes observed
below 500~cm$^{-1}$ agree with the group-theoretical predictions.
All the modes observed in the frequency ranges of the $\nu _{4}$,
$\nu _{2}$, and $\nu _{1}$ vibrations of BO$_{3}$ correspond well to
those derived using the correlation analysis and summarized in Fig.
\ref{fig9}. In contrast, there is one extra mode in the region of
the $\nu _{3}$ vibration. Possibly, the strong doublet at about
1220~cm$^{ - 1}$ in the (zx+zy)-polarization arises due to the Fermi
resonance \cite{her45} between the $\nu _{3}$ vibrations and an
overtone of the $\nu _{4}$ vibration.

The highest vibrational frequency (originating from the $\nu_3$
vibration of the BO$_{3}$ group) observed in
ErFe$_{3}$(BO$_{3})_{4}$ differs markedly from the highest
frequencies of both TO ($\vec {k}\parallel c$) and LO ($\vec
{k}\perp c$) phonons (1280 and 1414 cm-1, respectively, in
GdFe$_{3}$(BO$_{3})_{4}$). This is because in our particular non
oriented ErFe$_{3}$(BO$_{3})_{4}$  sample the direction of $\vec
{k}$ was arbitrary, while the frequencies of polar E modes depend
on the direction of $\vec {k}$.

Summarizing the RT results, it is concluded that (i) the number
and symmetries of the Raman modes observed for the
RFe$_{3}$(BO$_{3})_{4}$ in the high-temperature phase are fully
consistent with the predictions of group-theoretical analysis
based on the $R$32 space group; (ii) the modes above $\sim
$550~cm$^{ - 1}$ originate from the internal vibrations of the
BO$_{3}$ groups, while those below $\sim $550~cm$^{ - 1}$ can be
considered as external modes.

In the low-temperature $P$3$_{1}$21 phase, 20~$A_{1}$ (12 external
and 8 internal) and 40~$E$ (24 external an 16 internal) new modes
should appear, according to the results of the group-theoretical
analysis (see Table \ref{tab2}). These new modes arise due to two
reasons. First, the symmetries of the local positions for some of
the atoms reduce. The intensity of these new modes should be
proportional to the deviation from the former symmetric position.
Second, the primitive cell contains now not one but three formula
units, leading to an additional Davydov (factor-group) splitting
proportional to the strength of interaction between equivalent atoms
inside the new primitive cell. Both these effects are, as a rule,
small. That is why the number of the observed new modes
(10$A_{1}\oplus18E$ in GdFe$_{3}$(BO$_{3})_{4}$) is lower than the
number predicted by group-theoretical analysis. The same new modes
appear below $T_{s}$ in all the compounds RFe$_{3}$(BO$_{3})_{4}$,
with R=Gd, Tb, Er, and Y (see Fig.~5 and the bottom of Table III),
which strongly suggests the same $P$3$_{1}$21 low-temperature
structure in all of them.

\subsection{Weak first-order structural phase transition}
\begin{figure}
\includegraphics[width=\figwid]{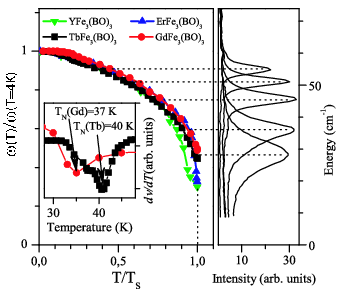}
\caption{\label{fig7} (Color online) The lowest-frequency mode
appears at the phase transition at a finite frequency
(characteristic of a first-order transition), but exhibits a
high-frequency shift with further lowering the temperature
(characteristic of a second-order transition). The inset shows the
derivative of the peak frequency in the vicinity of the magnetic
ordering temperature, evidencing the coupling of vibrational and
magnetic degrees of freedom (see text for discussion).}
\end{figure}

The most characteristic Raman feature that announces the phase
transition from the high-temperature $R$32 structure to the
low-temperature $P$3$_{1}$21 one is the sudden appearance of a
strong new low-frequency mode (see Figs.\ref{fig4}, \ref{fig6} and
\ref{fig7}) in the Raman spectra. As the intensities of new modes
are proportional to the squares of atomic displacements, ${I}\sim
\delta ^{2}$, the strongest new mode is believed to be associated
with the biggest displacements. A detailed analysis, according to
Ref. \cite{kli05}, of the structural changes shows that the
biggest displacements are those associated with the BO$_{3}$
"molecules". In particular, BO$_{3}$ triangles, perpendicular to
the $C_{3}$ axis in the $R$32 structure, tilt by $\sim $ 7$^{o}$
in the $P$3$_{1}$21 phase; the B1 atoms shift by $\sim $0.03~{\AA}
from the centers of regular triangles. The shifts of boron ions
relative to neighboring oxygen ions create local dipole moments;
their triangular arrangement corresponds to an antiferroelectric
ordering below the temperature of the structural phase transition.
This ordering manifests itself via a strong dielectric anomaly at
$T_{S}$ observed in our preliminary dielectric measurements (Fig.
\ref{fig10}).

The structural changes considered above give rise to many new
Raman active vibrational modes connected with the BO$_{3}$ groups
(see Table \ref{tab2}), in particular, to 4$A_{1}$ additional
librational modes. The energy value of the intense excitation
measured is in the typical range of the molecular librations. At 3
K, it is 56, 57, 72, and 76 cm$^{-1}$ for Gd(157), Tb(159),
Er(167), and Y(89) compounds, respectively. As the values for the
Er and Y compounds are very close, notwithstanding a big
difference in atomic masses of Er and Y, we conclude that the rare
earth does not take part in this lowest-frequency vibration. A
difference between the values for the compounds with relatively
big (Gd, Tb) and small (Er, Y) ions comes from the difference in
interatomic distances and, hence, force constants. The ensemble of
these observations strongly suggests that the phase transition in
RFe$_{3}$(BO$_{3})_{4}$ is correlated to a rotational mode of the
BO$_{3}$ groups.

The abrupt appearance of new phonon modes and the hysteresis
observed for their intensities when approaching $T_{S}$ from
different sides indicate the first-order character of the phase
transition. The energy of the low frequency mode strongly
increases upon lowering the temperature below $T_{S}$
(Fig.\ref{fig7}). At the lower temperature (3K) the frequency of
the mode has more than doubled. Such an anomalously large shift is
typical for soft modes that announce a second-order structural
phase transition. Apparently, we deal with a so-called "weak
first-order" phase transition. It is worth noting that the
lowering of the symmetry from $R$32 ($D_3^7 )$ to $P$3$_{1}$21
($D_3^4 )$ is indeed also compatible with a second-order phase
transition \cite{lan80}. Thus, the observed first-order character
of the phase transition does not depend on symmetry changes but
rather can arise from the third order therm in the Landau
expansion of the crystal free energy $h$, allowed by the crystal
symmetry, or from the negative sign of the fourth order
coefficient $u$:

\begin{equation}
\label{eq6} G = G_0 + rQ^2 + hQ^3 + uQ^4 +...,
\end{equation}
where Q is the order parameter \cite{bru81}.

Figure~\ref{fig11} shows the temperatures of the structural phase
transition in RFe$_{3}$(BO$_{3})_{4}$ plotted versus ionic radii of
R$^{3 + }$, as determined in Ref.~\cite{hin03} for powder samples
and in the present study for single crystals. A mismatch between our
data and those of Ref.~ \cite{hin03} could arise from a rather poor
quality of the Er and Y iron borate single crystals. A further study
is necessary to clarify this question.

Finally we note that NdFe$_{3}$(BO$_{3})_{4}$ preserves the $R$32
structure down to the lowest temperatures, while
RFe$_{3}$(BO$_{3})_{4}$ with R$^{3 + }$ smaller than Tb$^{3 + }$
have the low-temperature $P$3$_{1}$21 structure already at room
temperature.
\begin{figure}
\includegraphics[width=\figwid]{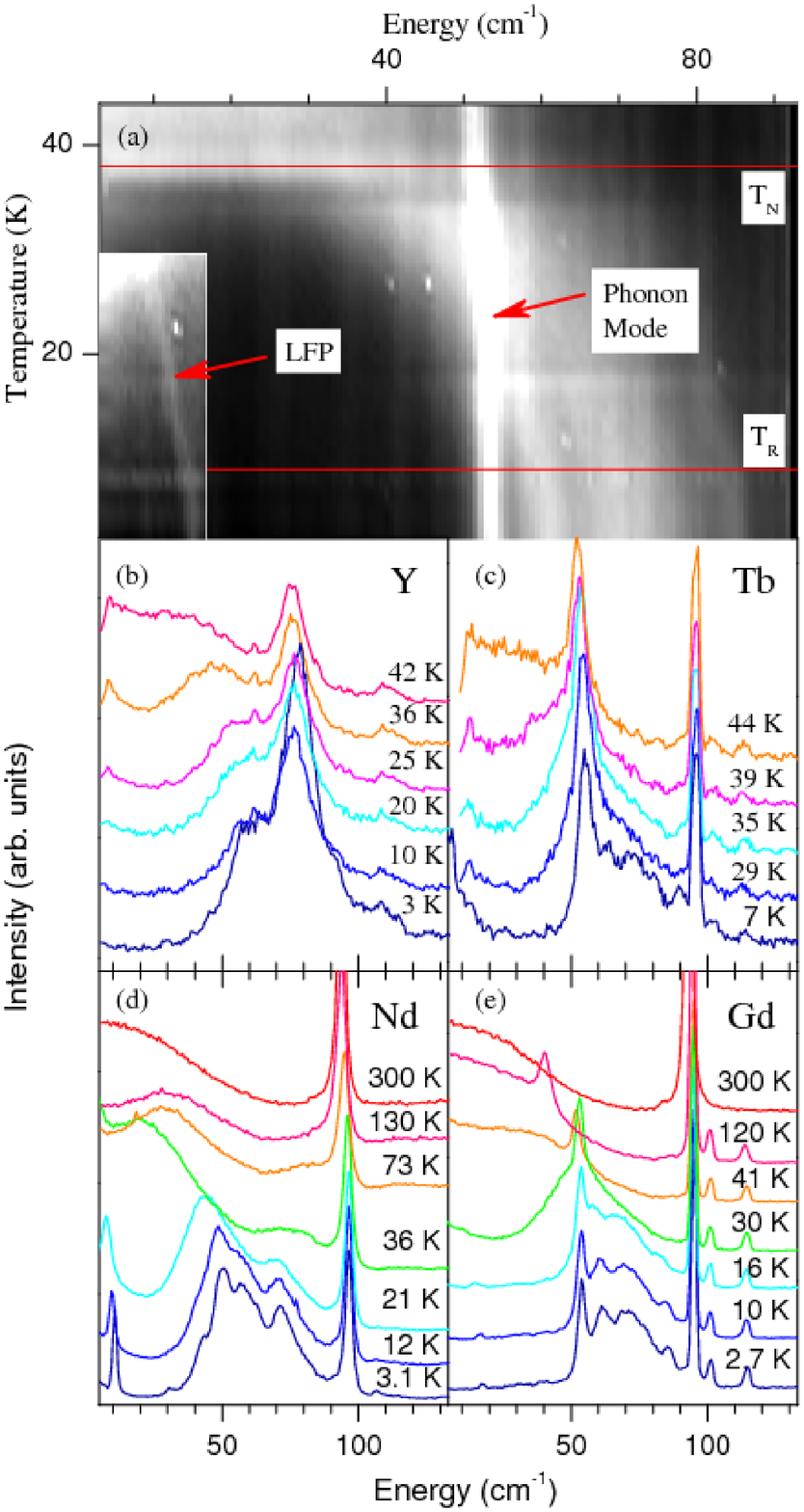}
\caption{\label{fig8} (Color online) Raman spectra of different
RFe$_{3}$(BO$_{3}$)$_{4}$ collected in (xz+yz) polarization. The
upper picture (a) shows the opening of a gap in the magnon Raman
scattering at $T_{N}$ = 37K for GdFe$_{3}$(BO$_{3}$)$_{4}$ (the
intensity scale of the left bottom corner of the picture is
optimized to evidence the low-frequency mode, and the dotted lines
are guides for evidencing the scattering features). The central
picture (b,c) show the magnetic scattering around $T_{N}$ for
YFe$_{3}$(BO$_{3}$)$_{4}$ and TbFe$_{3}$(BO$_{3}$)$_{4}$. The lower
one (d,e) shows the two-magnon spectra in the full temperature range
(2.7K$<T<$300K) for NdFe$_{3}$(BO$_{3}$)$_{4}$ on the left and for
GdFe$_{3}$(BO$_{3}$)$_{4}$ on the right. For the Gd sample a phonon
line is evident in proximity of the magnon feature. LFP indicates
the low-frequency peak, see the text for the discussion. Above
$T_{N}$ the paramagnon scattering survives up to at least nearly ten
times the ordering temperature. Different temperature measurements
in the panels b,c,d,e are depicted with an arbitrary offset for
clarity.}
\end{figure}

\subsection{Magnetoelastic coupling}
The inset of Fig.\ref{fig7} shows the derivative of the peak
frequency of the strongest new $A_{1}$ mode as a function of
temperature in the vicinity of the magnetic ordering temperature
$T_{N}$. The coupling of vibrational and magnetic degrees of
freedom is evident. The discontinuity in the phonon frequency
observed at $T=T_{N}$ is ascribed to magneto-elastic coupling. The
magnetic ordering causes spontaneous magnetostriction, that is,
atomic displacements. The latter, evidently, influence vibrational
frequencies. Recently, Ref.\cite{zve05} reported on the study of
magnetoelectric and magnetoelastic properties of
GdFe$_{3}$(BO$_{3})_{4}$. A strict correlation between both has
been shown experimentally and discussed theoretically. The authors
of Ref.\cite{zve05} assumed that, below $T_{N}$, the crystal
symmetry lowers, due to magnetoelastic coupling with spontaneous
magnetic moments lying in the ab-plane. This breaks the symmetry
of the antiferroelectric arrangement of the electric dipole
moments in the $P$3$_{1}$21 structure and leads to the appearance
of an electric polarization which could be responsible for the
weak growth of the dielectric constant in the temperature region
$T_{R}<T<T_{N}$ \cite{zve05}, (see also Fig. \ref{fig10}). The
proposed lowering of crystal symmetry below $T_{N}$, however,
should result in the appearance of new vibrational modes. No
indications for additional modes below $T_{N}$ have been found.
This does not necessarily mean that the model by Zvezdin
\textit{et al.} is in error. It might very well be that the
structural changes are simply too small to be detected by Raman
scattering.

\begin{figure}
\includegraphics[width=\figwid]{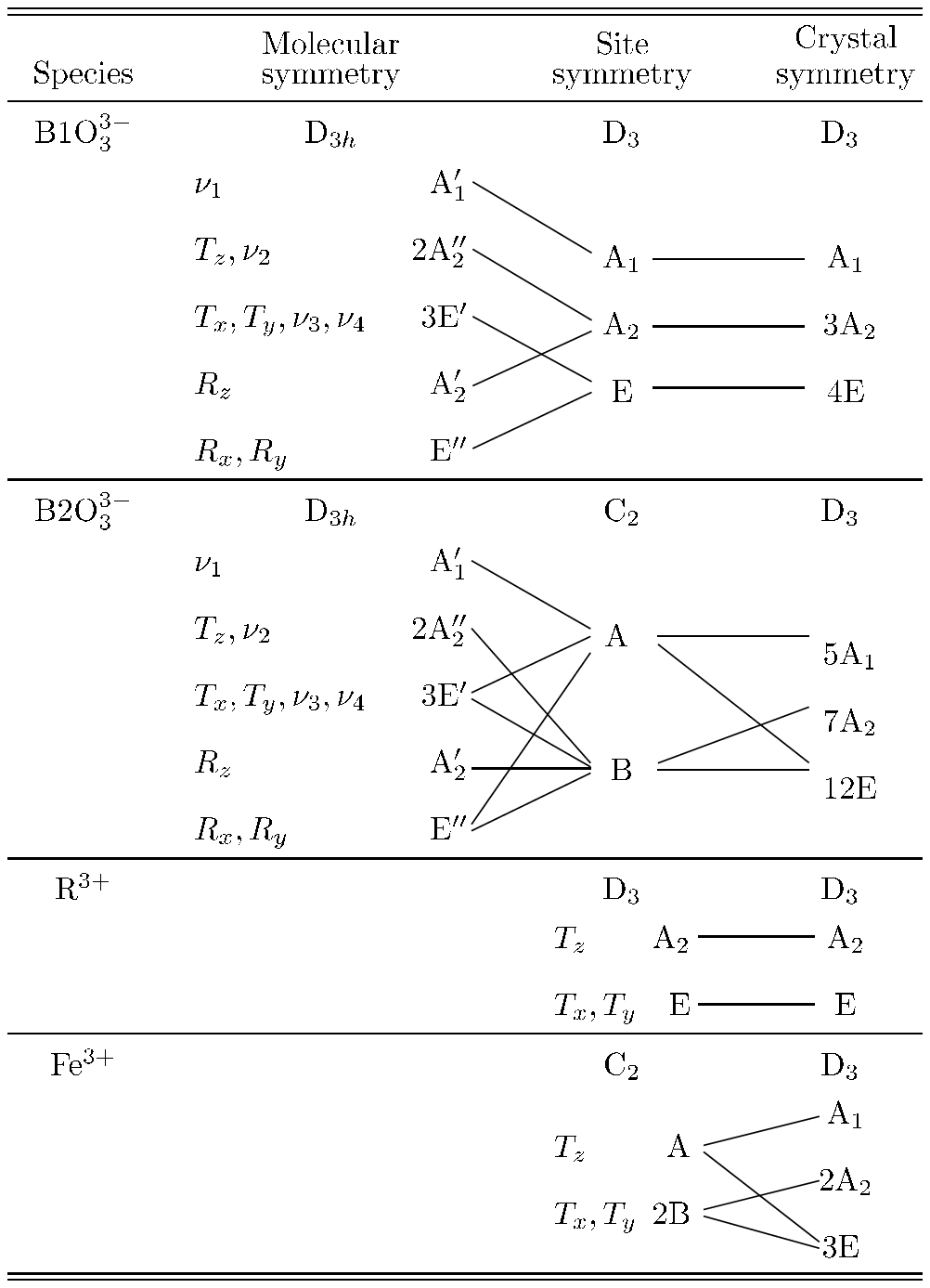}
\caption{\label{fig9} Correlation scheme for vibrational modes of
RFe$_{3}$(BO$_{3}$)$_{4}$, space group $R$32.}
\end{figure}

As Reference \cite{zve05} suggests, the spin-reorientation
first-order phase transition at $T_{R}$ into the antiferromagnetic
configuration of spins parallel to the c-axis recovers the
$P$3$_{1}$21 structure (that does not allow the presence of the
electric dipole moment) and, therefore, is expected to be
accompanied by a jump of the dielectric constant. As is clear from
Fig.~\ref{fig10}, the dielectric function indeed exhibits such a
jump, thereby strongly supporting the suggested scenario.

\subsection{Magnetic scattering}
As discussed in the introduction, due to the structural properties
of the RFe$_{3}$(BO$_{3})_{4}$ family, Fe atoms (S=5/2) are
arranged in helicoidal chains. As a result, the Fe-Fe distance
inside the chain is substantially smaller than the distance
between different chains. The main exchange interaction is
therefore expected to be between Fe ions on the same chain ($J_{ /
/ } \rangle \rangle J_ \bot )$, suggesting that low dimensionality
could play a crucial role in the magnetic properties of the
family. This, together with the presence of two coupled magnetic
sublattices makes the compounds with magnetic R$^{3 + }$ ions good
candidates for exotic magnetic ground states.

\begin{figure}
\includegraphics[width=\figwid]{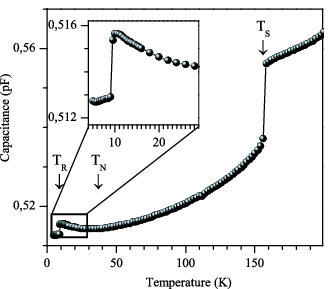}
\caption{\label{fig10} Dielectric measurements of
GdFe$_{3}$(BO$_{3}$)$_{4}$. The structural phase transition is made
evident by a step in the capacitance at $T_{S}$. No evidence is
found of the Fe ordering phase transition at $T_{N}$, while the
capacitance shows a discontinuity at the spin-reorientation phase
transition at $T_{R}$ (displayed in more detail in the Inset).}
\end{figure}

\begin{figure}
\includegraphics[width=\figwid]{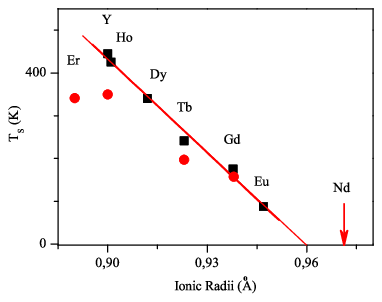}
\caption{\label{fig11} (Color online) Temperatures of the structural
phase transitions for different RFe$_{3}$(BO$_{3}$)$_{4}$ as a
function of the ionic radius. Our data on single crystals are shown
by circles, those of Ref. \cite{hin03} on powder samples are
presented by squares.}
\end{figure}

A complex magnetic behavior has indeed been found and widely
discussed in GdFe$_{3}$(BO$_{3})_{4}$. The first magnetization
measurements on oriented single crystals were interpreted under
the assumption that both Fe and Gd subsystems order
antiferromagnetically at $T_{N}$=40 K~\cite{bal03}. In this model,
below 10~K, the total magnetic moment of the three Fe sublattices
(S~=~5/2) oriented at polar angles $\sim $ 60\r{ } to the $c$ axis
is compensated by the Gd moment (S~=~7/2) oriented along the $c$
axis. Sharp singularities in magnetization in the magnetic field
H$\vert \vert  c$ observed below 10 K were attributed to spin-flop
transitions in the Fe subsystem. It was assumed further that in
the range of temperatures between 10 and~40 K the magnetic moments
of every Fe sublattice fall into the \textit{ab} plane,
perpendicular to the $c$ axis, preserving the 120\r{ } azimuthal
orientation.

This model was later reconsidered in Ref.~\cite{lev04}. It has been
suggested, on the basis of the analysis of the Nd-probe spectra and
of the specific heat data, that the Gd subsystem does not undergo a
magnetic phase transition at $T_{N}$ but only gets polarized by the
Fe subsystem. The following model has been put forward. At
$T_{N}$=37~K, the iron magnetic moments order antiferromagnetically
in the direction perpendicular to the c-axis and polarize the Gd
subsystem. Below $T_{R}$=9~K, the magnetic moments order
antiferromagnetically along the c-axis. The temperature $T_{R}$=9~K
corresponds to a first-order spin-reorientation magnetic phase
transition \cite{lev04}. Simultaneously and independently, the same
conclusion has been drawn in Ref.\cite{pan04} where magnetic
structures and magnetic phase transitions in
GdFe$_{3}$(BO$_{3})_{4}$ were studied using antiferromagnetic
resonance experiments. This study revealed also a detailed picture
of the magnetic structure and anisotropy for different magnetic
phases in GdFe$_{3}$(BO$_{3})_{4}$ as a function of both temperature
and magnetic field. At $T_{N}$=38~K, the Fe magnetic subsystem
orders into a two-sublattice collinear easy-plane antiferromagnet
and polarizes the Gd spins, which also form a two-sublattice
antiferromagnetic subsystem. The anisotropy constant of the Gd
subsystem has an opposite sign to that of the Fe subsystem. The Gd
contribution to the total anisotropy energy grows with lowering the
temperature, in conjunction with the growing polarization of the Gd
subsystem, and becomes appreciable below $\sim $20~K. At
$T_{R}$=10~K the total energy of anisotropy changes its sign which
results in the spontaneous spin reorientation transition.

The long-range order of the spins on the iron subsystem manifests
itself in the inelastic light scattering experiments with two main
features. The first one, as reported in Fig.\ref{fig8}, is the
arising of a low-frequency peak (LFP) at $\sim $10~cm$^{ - 1}$ in
NdFe$_{3}$(BO$_{3})_{4}$ and at $\sim $18~cm$^{ - 1}$ in
GdFe$_{3}$(BO$_{3})_{4}$. These peaks soften and broaden upon
approaching $T_{N}$. The energy of 10~cm$^{ - 1}$ in
NdFe$_{3}$(BO$_{3})_{4}$ has been identified, by absorption
spectroscopy of Nd$^{3 + }$ crystal-field levels, as the exchange
splitting energy of the Nd$^{3 + }$ ground Kramers doublet at 5 K
arising from the interaction with the ordered Fe sublattice
\cite{chu04}. The temperature dependences of the frequency and the
linewidth of the Raman LFP are the same as those of the ground-state
splitting and, respectively, the ground-level width found from
optical spectroscopy \cite{chu04}. Therefore the 10~cm$^{ - 1}$
Raman mode observed in NdFe$_{3}$(BO$_{3})_{4}$ is assigned to spin
flip scattering on a single Nd$^{3 + }$ moment in the effective
field created by the Fe sublattice. Most likely, the 18~cm$^{ - 1}$
excitation observed in GdFe$_{3}$(BO$_{3})_{4}$ has the same origin.
The absence of LFP in Raman scattering of \yfb ~with nonmagnetic
Y$^{3+}$ (see Fig. \ref{fig8}) is in favor of such interpretation.
The crystal field of \rfb ~splits the ground-state multiplet of an
ion R$^{3+}$ with odd number of electrons into $\Gamma_{4}$ and
$\Gamma_{56}$ Kramers doublets within the D$_3$ site symmetry. The
analysis of Raman selection rules for the ground-state spin flip
scattering shows that in the case of the $\Gamma_{4}$ ground state
all the components of the Raman tensor are allowed while in the case
of the $\Gamma_{56}$ ground state only the antisymmetric xy-yx
component is Raman active. Thus, a relatively strong (xz) Raman LFP
of \nfb~ points to the $\Gamma_{4}$ symmetry of the Nd$^{3+}$ ground
state.

The second major feature of the iron spin ordering is the arising
of a broad structured scattering band around 60~cm$^{ - 1}$
ascribed to two-magnon Raman scattering involving the creation of
a pair of magnons with wave vectors $\vec{k}$ and -$\vec{k}$.
Below the N\'{e}el temperature this is a characteristic feature
for all the different compound investigated of the
RFe$_{3}$(BO$_{3})_{4}$ family (R=Y, Er, Tb, Gd and Nd).  Figure
\ref{fig8} shows the temperature evolution of this broad signature
of magnetic scattering. The spectra reported in Fig.\ref {fig8}
are found in the (xz+yz) polarization only. No evidence of
magnetic scattering is found when the $\vec{k}$ vector of light is
parallel to the c-axis (z(xx+yy+xy)z).

\begin{figure}
\includegraphics[width=\figwid]{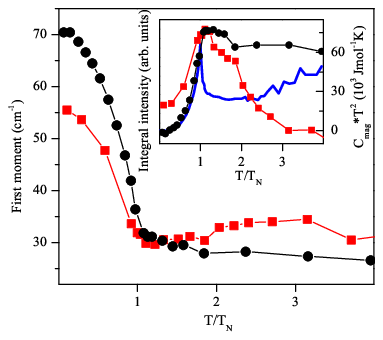}
\caption{\label{fig12} (Color online) First moment $<\omega>$ of the
magnetic excitation vs temperature calculated from 5 to 85~cm$^{-1}$
for GdFe$_{3}$(BO$_{3}$)$_{4}$ (circles) and
NdFe$_{3}$(BO$_{3}$)$_{4}$ (squares). The inset shows the integrated
intensity of the magnetic signal at different temperature, compared
to the magnetic heat capacity of YFe$_{3}$(BO$_{3}$)$_{4}$ (solid
line)\cite{hin03}.}
\end{figure}

As discussed in the previous section, NdFe$_{3}$(BO$_{3})_{4}$
does not undergo a structural phase transition and the crystal
space group remains the high temperature one ($R$32), while the
symmetry of all the other compounds investigated (for R=Gd, Tb, Y,
and Er) is reduced to the space group $P$3$_{1}$21. It is evident
from Fig.\ref{fig8} that the magnetic scattering spectra of all
iron borates are quite similar. So, the presence of inequivalent
Fe chains \cite{kli05} in the low-temperature structure of
RFe$_{3}$(BO$_{3})_{4}$, R~=~Gd, Tb, Y, and Er does not strongly
affect the magnetic excitation spectra, and the spectra can be
analyzed taking into account only the four different magnetic ions
present in the $R$32 unit cell (three Fe$^{3 + }$ ions and one
R$^{3 + }$ ion). Still, a rather complicated two-magnon spectrum
is expected, strongly depending on the anisotropy parameters. At
the lowest temperature reached (T=2.7 K) the two-magnon spectra
indeed show a complex structure exhibiting at least three main
peaks (Fig.\ref{fig8}). As the most efficient mechanism of
two-magnon scattering in antiferromagnets is usually the
exchange-scattering mechanism, and the strongest exchange
interaction is between the iron atoms along the helicoidal chains,
these peaks presumably arise from three magnon branches
representing spin excitations on the iron chains.

The broad magnetic scattering feature of Gd, Tb, Er and Y
compounds is approximately centered around $\simeq$70~cm$^{ - 1}$,
while the one of Nd is centered at lower frequency
($\simeq$60~cm$^{ - 1}$). This suggests that the bigger ionic
radius of Nd$^{3 + }$, causes the interatomic Fe-Fe distances to
be larger, and therefore the exchange interaction to be smaller,
this would be consistent with the scaling of $T_{N}$ with the rare
earth's ionic radii \cite{hin03}. Unfortunately, due to the
presence of phonon modes at the same frequency of the magnetic
excitation, it is not possible to determine the frequency scaling
of the magnetic scattering among the other compounds (Gd, Tb, Er
and Y) to confirm it.

At the temperature of the spin-reorientational transition
($T_{R})$, no drastic changes are observed in the magnon spectra.
This observation confirms that the observed scattering is mainly
due to magnetic excitations on the Fe$^{3 + }$ chains, and sheds
light on the proposed exotic magnetic ordering of Fe spins for the
GdFe$_{3}$(BO$_{3})_{4}$ compound. The configuration with
120$^\circ$ angle between nearest neighbor Fe$^{3 + }$ spins for
$T_{N}>T>T_{R}$ which then reorient at $T_{R}$ into an easy axis
antiferromagnet along the c-axis as proposed in \cite{bal03} is
not consistent with our observation. Such a drastic change in the
spin configuration would induce a major change in the two-magnon
dispersion and inelastic scattering spectra. The picture of a
reorientational phase transition at $T_{R}$, as proposed in
\cite{pan04,lev04}, survives assuming an easy-axis/easy-plane
anisotropy perpendicular to the c-axis in the temperature region
$T_{R}<T<T_{N}$. Indeed, considering a simple Heisenberg-type
Hamiltonian, where the interaction depends only on the nearest
neighbor scalar product, an easy axis anisotropy perpendicular or
parallel to the c-axis would produce similar magnetic excitation
spectra and therefore similar two-magnon spectra.

A striking feature of the magnetic scattering, as shown in Fig.
\ref{fig8}, is the persistence of the magnetic scattering in an
anomalous shape up to extremely high temperature above $T_{N}$. In
the three dimensional antiferromagnet NiF$_{2}$ this ``paramagnon
scattering'' has been observed well into the paramagnetic phase up
to 4$T_{N}$ \cite{fle69}. The first moment of the frequency
$<\omega>$ as a function of temperature in NiF$_{2}$ and MnF$_{2}$
exhibited a fast decrease to lower frequency ($\simeq10\%$ of the
''0 temperature value") at $T_{N}$ and a subsequent smooth
decrease to zero for T $>$ $T_{N}$, in excellent qualitative
agreement with the theory prediction for the moments \cite{cot86}.
Also in the case of RFe$_{3}$(BO$_{3})_{4}$ the two-magnon
spectrum softens continuously its frequency approaching the
N\'{e}el temperature (Fig.\ref{fig8}). Yet, surprisingly, the
paramagnon scattering survives at all measured temperatures above
$T_{N}$ ($i.e.$ up to about 10 times $T_{N})$. As reported in Fig.
\ref{fig12}, the first moment of the excitation decreases at
$T_{N}$ less than expected and approaches zero only very slowly
above $T_{N}$. Considering NdFe$_{3}$(BO$_{3})_{4}$, the minimum
at $T_{N}$ and subsequent increases with temperature up to
2.5~$T_{N}$ of the first moment is possibly due to the presence of
luminescence lines overlapping with features of the magnetic
scattering. This is likely also the origin of the temperature
shift to high frequency (lower absolute energy) of the excitation
observed at around 80cm$^{-1}$.

\begin{figure}
\includegraphics[width=\figwid]{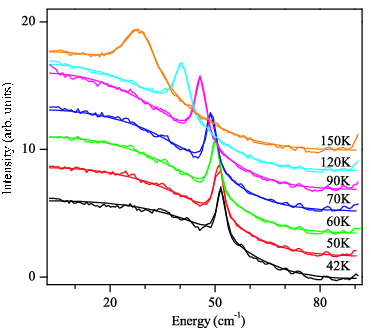} \caption{(Color
online) Low frequency Raman scattering above the N\'eel temperature
for GdFe$_{3}$(BO$_{3})_{4}$. The different temperature are depicted
with an offset for clarity. The paramagnon scattering is present up
to almost $4T_{N}$. The fit is obtained with an fenomenological
curve $f(\omega)=\frac{G}{1+aG}+L$, where $G$ is a Gaussian
distribution centered at 0 frequency, and $L$ is a Lorenzian
distribution to take into account the presence of the vibrational
mode (see text for discussion).} \label{fig13}
\end{figure}

From a theoretical point of view the presence of quasi-elastic
scattering well into the paramagnetic phase of three-dimensional
antiferromagnets has been ascribed to two possible origins,
namely, spin diffusion \cite{rich74} and spin density fluctuations
or energy diffusion \cite{hall78}. In the scenario of spin
diffusion, absent in a perfect one-dimensional system, the
four-spin correlation function (describing the Raman response) is
well approximated by a Gaussian shape centered at zero energy. In
contrast, for the energy diffusion scenario one expects a
lorentzian lineshape in the absence of strong spin-lattice
interactions. The integrated intensity $I_m$ in the energy
diffusion scenario is expected to follow the magnetic contribution
to the heat capacity $C_m$ as $I_m\propto C_m
T^2$\cite{rich74,hall78}. In the presence of strong spin-lattice
coupling this later relation is expected to remain valid, even
though the low energy scattering is expected to broaden
significantly due to the rapid relaxation of the magnetic
excitation energy into the lattice.

The shape of the low-energy scattering feature in the paramagnetic
phase is different from any previously reported. Neither a
Lorenzian nor a Gaussian distribution can fit our data. This
anomalous shape of the paramagnon scattering continuum is a common
feature of all the compounds investigated; Fig.\ref{fig13} depicts
the observed spectra of GdFe$_{3}$(BO$_{3})_{4}$ as an example.
The observation that the spectra can not be modelled with either a
Gaussian nor a Lorentzian response might evidence the presence of
important spin-lattice interactions, in line with the observations
discussed in the previous section. A comparison of the integrated
intensity for \gfb\ and \nfb\ to the magnetic heat capacity
determined by Hinatsu {\em et al.}\cite{hin03} (see inset fig. 12)
for \yfb\ does not show a fair agreement in the `paramagnetic'
phase, even though there is a good correlation below the phase
transition. Apparently energy diffusion, combined with
spin-lattice coupling does not explain the observed scattering. On
the other hand, as shown in Figure \ref{fig13} the spectra for
GdFe$_{3}$(BO$_{3})_{4}$ can be fitted nicely at all temperature
with a broad phenomenological function consisting of a
renormalized Gaussian distribution
$I(\omega)=\frac{G(\omega)}{1+\alpha G(\omega)}+L$, where
$G(\omega)$ is a pure Gaussian, and $L(\omega)$ is a lorentzian to
account for the phonon line observed in the spectra. This equation
is reminescent of the response function of interacting magnons in
low dimensional quantum spin systems, although it would be unclear
why this would still hold for a $S=5/2$ system at high
temperatures. At present one can not draw a definite conclusion,
but the results do strongly suggest that short range order
correlations in the spin system survives up to quite high
temperatures. Most likely an interplay between the low
dimensionality and strong spin-lattice interactions are
responsible for the observed unique behavior of the magnetic
scattering up to several times $T_{N }$, and deserves further
studies.

\section{Summary}
In summary, structural, magnetic, and magneto-elastic properties
of various members of the RFe$_{3}$(BO$_{3})_{4}$ family (R~=~Gd,
Tb, Nd, Er, and Y) have been studied using primarily inelastic
light scattering. The compounds show a weak first-order structural
phase transition which results in the activation of a strong mode
of BO$_{3}$ librational nature, and its subsequent temperature
evolution. The scattering spectra observed in the low-temperature
phase are fully consistent with the low-temperature structure
reported earlier. A detailed analysis of the vibrational spectra
at the temperature of the magnetic ordering transition showed a
strong magneto-elastic coupling for all the compounds. The
analysis of the two-magnon Raman scattering in the magnetic phases
showed qualitatively similar scattering spectra for the five
compounds investigated, evidencing that the structural differences
between them do not strongly affect the magnetism and the magnetic
excitation spectra. Finally an unprecedented an intriguing
paramagnon scattering continuum up to quite high temperatures. The
origin of this scattering does not seem to be the usual energy
diffusion observed in a variety of other magnetic system, and
seems to indicate the presence of short-range order spin-spin
correlations arising from the low-dimensional nature of these
compounds.

\begin{acknowledgments}
The authors are grateful to M. Mostovoy and D. Khomskii for
valuable discussions. This work was partially supported by the
Stichting voor Fundamenteel Onderzoek der Materie [FOM,
financially supported by the Nederlandse Organisatie voor
Wetenschappelijk Onderzoek (NWO)]. SAK, MNP, and LNB acknowledge
the support of the Russian Foundation for Basic Research, grants ¹
04-02-17346 and ¹ 03-02-16286, and the Russian Academy of Sciences
under the Programs for Basic Research.
\end{acknowledgments}

\begin{table*}
  \caption{\label{tab1} Frequency range and assignment of the vibrations of the \booo-group.
    The second column lists the assignments for the free ions.
    The third column shows the expected internal modes and their symmetries in the high temperature
    phase of the rare-earth iron borates (the superscripts 1 and 2 refer to the different borate groups
    in the rare-earth iron borates).}
  \begin{ruledtabular}
    \begin{tabular}{lll}
        Frequency   range (\cm)& \booo\ vibrations (D$_{3\mathrm{h}}$) &    Crystal vibrations\\\hline
                600 & $\nu_4(E')$: in-plane bending & $\nu_4^{(2)}(A_1)+\nu_4^{(1)}(E)+2\nu_4^{(2)}(E)$ \\
                700-800 & $\nu_2(A''_2)$    & $\nu_2^{(2)}(E)$ \\
                950 & $\nu_1(A'_1)$: symmetric breathing    & $\nu_1^{(1)}(A_1)+\nu_1^{(2)}(A_1)+\nu_1^{(2)}(E)$ \\
                1250-1400 & $\nu_3(E')$: asymmetric breathing   & $\nu_3^{(2)}(A_1)+\nu_3^{(1)}(E)+\nu_3^{(2)}(E)$
    \end{tabular}
  \end{ruledtabular}
\end{table*}

\begin{table*}
  \caption{
    The origin of vibrations for two structural modifications of \rfb\ and the new modes that
    should appear below the temperature of the structural phase transition $T_s$.
    \label{tab2}
  }
  \begin{ruledtabular}
    \begin{tabular}{llll}
       species & $T>T_s$: $R$32 ($D^7_3$) & $T<T_s$:  $P$3$_1$21 ($D^4_3$) & New modes \\\hline
       R    & $A_2 \oplus E$ &  $A_1 \oplus 2A_2 \oplus 3E$&    $A_1\oplus 2E\oplus A_2$ \\
       Fe   & $A_1 \oplus 2A_2 \oplus 3E$ & $4A_1 \oplus 5A_2 \oplus 9E$ & $3A_1 \oplus 6E\oplus 3A_2$\\
       external \booo & & & \\
       \hspace*{5mm} translational &    $A_1 \oplus 3A_2 \oplus 4E$
                    &   $5A_1 \oplus 7A_2 \oplus 12E$ & $4A_1 \oplus 8E\oplus 4A_2$\\
       \hspace*{5mm} librational & $A_1 \oplus 3A_2 \oplus 4E$
                    &   $5A_1 \oplus 7A_2 \oplus 12E$ & $4A_1 \oplus 8E\oplus 4A_2$\\
       internal \booo & $4A_1 \oplus 4A_2 \oplus 8E$ & $12A_1 \oplus 12A_2 \oplus 24E$
                    & $8A_1 \oplus 16E\oplus 8A_2$\\\hline
       Total & $7A_1 \oplus 13A_2 \oplus 20E$ & $27A_1 \oplus 33A_2 \oplus 60E$ & $20A_1 \oplus 40E\oplus 20A_2$
    \end{tabular}
  \end{ruledtabular}
\end{table*}

\begin{table*}
  \caption{
    Observed vibrational modes for the high-temperature phase of several rare-earth iron borates.
    \label{tab3}
  }
  \begin{ruledtabular}
    \begin{tabular}{llllllllllll}
    \multicolumn{12}{l}{External modes ($3A_1 \oplus 11E$)}\\
\hline
    \multicolumn{3}{l}{\nfb}&\multicolumn{3}{l}{\gfb}&\multicolumn{2}{l}{\tfb}&\multicolumn{2}{l}{\efb}&\multicolumn{2}{l}{\yfb}\\
    $A_1$&  $E_{TO}$    & $E_{LO}$ & $A_1$ &    $E_{TO}$ & $E_{LO}$ &   $A_1$ & $E$ &   $A_1$ & $E$ &   $A_1$ & $E$\\
    180& 89& 93&180& 84& 93&180& 93&181&84&180&107\\
    298&   &159&307&160&160&309&159&312&160&312&158\\
        473&   &193&475&195&198&476&198&472&199&468&200\\
             &   &232&   &   &232&   &230&   &225& & \\
             &260&266&   &270&270&   &270&   &272& & 268\\
             &272&281&   &273&287&   &287&   & & & \\
             &312&332&   &315&330&   &330&   &325& & 327\\
             &354&356&   &352&357&   &355&   &350&&354\\
             &   &384&   &391&391&   &392&   &395&&390\\
             &   &439&   &443&443&   &444&   &442&&441\\
             &475&488&   &   &488&   &480&   &475&&470\\
\hline\hline
\multicolumn{12}{l}{Internal modes ($4A_1 \oplus 8E$)}\\
\hline
    \multicolumn{3}{l}{\nfb}&\multicolumn{3}{l}{\gfb}&\multicolumn{2}{l}{\tfb}&\multicolumn{2}{l}{\efb}&\multicolumn{2}{l}{\yfb}\\
    $A_1$&  \multicolumn{2}{l}{$E$} & $A_1$ &   $E_{TO}$ & $E_{LO}$ &   $A_1$ & $E$ &   $A_1$ & $E$ &   $A_1$ & $E$\\
 636&\multicolumn{2}{l}{ 579}& 638&    & 580& 635& 579& 632& 579&631&576\\
 950&\multicolumn{2}{l}{ 625}& 957& 631& 633& 957& 632& 960& 632&&\\
 990&\multicolumn{2}{l}{ 669}& 990& 670& 676& 988& 676& 988& 675&984&672\\
1220&\multicolumn{2}{l}{ 734}&1230&    & 735&1220& 734&1230& 730&&732\\
        &\multicolumn{2}{l}{ 968}&    &    & 968&    & 966&    & 960&&959\\
        &\multicolumn{2}{l}{1195}&    &    &    &    &    &    &    &&1219\\
        &\multicolumn{2}{l}{1218}&    &    &1212&    &1214&    &1220&&1230\\
        &\multicolumn{2}{l}{1244}&    &    &1229&    &1230&    &1233&&1408\\
        &\multicolumn{2}{l}{1260}&    &1198&1250&    &1254&    &    &&\\
        &\multicolumn{2}{l}{1413}&    &1280&1414&    &1411&    &1342&&\\

\hline\hline \multicolumn{12}{l}{Modes appearing in the low T phase:  $20A_{1}$ (12 ext. and 8 int.) $\oplus$ $40E$ (24 ext. an 16 int.)}\\
\hline
    \multicolumn{3}{l}{\nfb}&\multicolumn{3}{l}{\gfb}&\multicolumn{2}{l}{\tfb}&\multicolumn{2}{l}{\efb}&\multicolumn{2}{l}{\yfb}\\
    \multicolumn{3}{c}{-} & $A_1$ &   \multicolumn{2}{l}{$E$} &   $A_1$ & $E$ &   $A_1$ & $E$ &   $A_1$ & $E$\\
\multicolumn{3}{c}{-}  & 53 &\multicolumn{2}{l}{101}& 54 &       &72  &    &76  &\\
\multicolumn{3}{c}{-}  &144 &\multicolumn{2}{l}{114}&    &       &149 &    &150 &111 \\
\multicolumn{3}{c}{-}  &203 &\multicolumn{2}{l}{167}&    & 169   &202 & 170&206 &173 \\
\multicolumn{3}{c}{-}  &233 &\multicolumn{2}{l}{206}&232 & 207   &234 & 208&    &211 \\
\multicolumn{3}{c}{-}  &244 &\multicolumn{2}{l}{254}&247 & 256   &    & 250&246 &253 \\
\multicolumn{3}{c}{-}  &263 &\multicolumn{2}{l}{276}&265 & 278   &    & 278&    &280 \\
\multicolumn{3}{c}{-}  &    &\multicolumn{2}{l}{281}&    & 282   &    & 284&    &287 \\
\multicolumn{3}{c}{-}  &    &\multicolumn{2}{l}{305}&    & 306   &    & 305&    &307 \\
\multicolumn{3}{c}{-}  &368 &\multicolumn{2}{l}{310}&368 & 311   &    &    &    &    \\
\multicolumn{3}{c}{-}  &385 &\multicolumn{2}{l}{337}&    & 337   &    &    &    &335 \\
\multicolumn{3}{c}{-}  &    &\multicolumn{2}{l}{374}&    & 375   &    & 373&    &370 \\
\multicolumn{3}{c}{-}  &    &\multicolumn{2}{l}{   }&    &       &    &    &    &378 \\
\multicolumn{3}{c}{-}  &    &\multicolumn{2}{l}{395}&    & 398   &    &    &    &\\
\multicolumn{3}{c}{-}  &677 &\multicolumn{2}{l}{472}&679 & 473   &    & 470&    &\\
\multicolumn{3}{c}{-}  &724 &\multicolumn{2}{l}{596}&723 & 597   &    &    &    &\\
\multicolumn{3}{c}{-}  &    &\multicolumn{2}{l}{654}&953 & 654   &    &    &    &\\
\multicolumn{3}{c}{-}  &    &\multicolumn{2}{l}{669}&    & 670   &    &    &    &\\
\multicolumn{3}{c}{-}  &    &\multicolumn{2}{l}{955}&    & 954   &    &    &    &\\
\multicolumn{3}{c}{-}  &    &\multicolumn{2}{l}{968}&    & 966   &    &    &    &\\
    \end{tabular}
  \end{ruledtabular}
\end{table*}

\end{document}